\documentclass[10pt,journal,twocolumn]{IEEEtran}
\usepackage{subfigure}
\usepackage{enumerate}

\usepackage{amsfonts}
\usepackage{amsmath}
\usepackage{amssymb}
\usepackage{amsthm}
\usepackage{cite}
\usepackage{color, soul}
\usepackage{bm}
\usepackage{array}
\usepackage{multirow}
\usepackage{booktabs}
\usepackage[pdftex]{graphicx}
\usepackage{footnote}

\usepackage{epsfig}

\begin{document}

\title{Follow Whom? Chinese Users Have Different Choice}

\author{Zhaoqun Chen, Pengfei Liu, Xiaohan Wang, and Yuantao Gu %
\thanks{Z. Chen, P. Liu, X. Wang, and Y. Gu are with the Department of Electronic Engineering, Tsinghua University, Beijing 10084, China. The corresponding author for this paper is Yuantao Gu (email: gyt@tsinghua.edu.cn)}}
\maketitle

\begin{abstract}
Sina Weibo, which was launched in 2009, is the most popular Chinese micro-blogging service. It has been reported that Sina Weibo has more than 400 million registered users by the end of the third quarter in 2012. Sina Weibo and Twitter have a lot in common, however, in terms of the following preference, Sina Weibo users, most of whom are Chinese, behave differently compared with those of Twitter.

This work is based on a data set of Sina Weibo which contains 80.8 million users' profiles and 7.2 billion relations and a large data set of Twitter. Firstly some basic features of Sina Weibo and Twitter are analyzed such as degree and activeness distribution, correlation between degree and activeness, and the degree of separation. Then the following preference is investigated by studying the assortative mixing, friend similarities, following distribution, edge balance ratio, and ranking correlation, where edge balance ratio is newly proposed to measure balance property of graphs. It is found that Sina Weibo has a lower reciprocity rate, more positive balanced relations and is more disassortative. Coinciding with Asian traditional culture, the following preference of Sina Weibo users is more concentrated and hierarchical: they are more likely to follow people at higher or the same social levels and less likely to follow people lower than themselves. In contrast, the same kind of following preference is weaker in Twitter. Twitter users are open as they follow people from levels, which accords with its global characteristic and the prevalence of western civilization. The message forwarding behavior is studied by displaying the propagation levels, delays, and critical users. The following preference derives from not only the usage habits but also underlying reasons such as personalities and social moralities that is worthy of future research. To the best of our knowledge, this is the first comparative work focusing on the following behavior using both large-scale data set of a global and a Chinese local online social networks.

\end{abstract}




\section{Introduction}
Twitter, a world-wide popular online microblogging service has reached a commercial success and attracted many researchers' attention. Due to some reasons, Twitter is hard to be accessed in mainland China. Many Chinese local micro-blogging services sprang up around 2009 including Sina Weibo, Tencent Weibo, and Sohu Weibo. Sina Weibo is the most popular one with more than 400 million users by the end of the third quarter in 2012. ``Weibo'' is the Chinese word for ``microblog''. Sina Weibo and Twitter share some basic features. One user can ``follow'' any other user to become his or her follower without any verification or approval. Users can post short messages (called tweets for Twitter and weibos for Sina Weibo) within certain length on their main pages and then all their followers will receive the messages. Tweets contain only text but weibos allow pictures and videos to be attached. Messages can be forwarded or ``retweet''. This mechanism enhances the power of message propagation and one message is able to cover a very large range in a very short time. Although Sina Weibo has already opened up its overseas markets to users from other countries, almost all the Sina Weibo users now are Chinese, while Twitter users are globally distributed except certain countries block it. Such difference in constituency results in the macro differences between Sina Weibo and Twitter.

Researches have shown that distinguishes from other online social networks like Facebook and MySpace, the unique unidirectional ``follow'' mechanism makes twitter has another role of social media. Because people use Twitter not only to follow their friends but also online celebrities and organizations in order to get news and gossips. The network structures of Sina Weibo and Twitter are very complex. However, the features of network structure at macro level are directly caused by every user's following preference at micro level. It becomes essential to figure out whether all the users have the similar following preference and how users' attributes such as nationalities and cultural backgrounds influence the choice about following whom. This is the main purpose of this paper.

In this work, large-scale data set is used to comparatively analyze both Sina Weibo and Twitter. The data set of Sina Weibo contains 80.8 million users' profiles and 7.2 billion relations that covers about $20\%$ number of all users. The data set of Twitter is from \cite{koreatwitter} and it contains 41 million users and 1.5 billion relations. Firstly some basic features of Sina Weibo and Twitter are analyzed such as degree and activeness distribution, correlation between degree and activeness, and the degree of separation. Then the following preference is investigated by studying the assortative mixing, friends' similarity, following distribution, and edge balance ratio, where edge balance ratio is newly proposed to measure a graph's balance property \cite{edgebalanceratio}. Sina Weibo and Twitter users are ranked by the number of followers and PageRank. The ranking correlation reflects the diversity of following preference. Based on the following preference and the social media role of micro-blogging services, the message forwarding behavior is studied by displaying the propagation levels, delays, and critical users. The contribution of our work reveals the difference of following preference between Sina Weibo and Twitter. To the best of our knowledge, this is the first comparative work focusing on the following behavior using both large-scale data set of a global and a Chinese local online social networks.

The rest of this paper is organized as follows. The related works are briefly reviewed in Section \ref{chap:relatedwork}. In Section \ref{chap:basic} some basic analysis is performed on Sina Weibo and Twitter. The main contribution of this work is contained in Section \ref{chap:followingpreference}, where the following preference is compared between Twitter users and Sina Weibo users. In Section \ref{chap:ranking}, users are ranked by the number of followers and PageRank. In Section \ref{chap:propagation}, two real examples are displayed. The propagation level, delay, coverage, pattern, and the important users in the process are studied. The conclusion goes in Section \ref{chap:conclusion}.

\section{Related works}
\label{chap:relatedwork}

Twitter has been studied widely. Java et al. studied the topological and geographical properties of Twitter \cite{javawhywetwitter}. Kwak et al. presented a quantitative study on the entire Twittersphere and information diffusion \cite{koreatwitter}. They used basic statistical methods to analyze topological features and found non-power-law distribution, a short effective diameter, and low reciprocity for Twitter, which indicates the differences from human social networks. Then it is concluded that Twitter is a news media more than a social network. Krishnamurthy et al. presented a detailed characterization of Twitter and studied user behaviors including following, geographic growth patterns, and other aspects \cite{fewchirps}.  Through comparing three different measures of influence: in-degree, retweets, and mentions, Cha et al. found the number of followers is not related to the number of retweets and mentions \cite{chamillionfollower}. Yang et al. studied the hashtag used in tweets propagation and reported the hashtag in Twitter play a dual role, a bookmark of content and the symbol of a community membership \cite{yanghashtag}. Kitsak found most efficient spreaders are those located within the core of the network by the k-shell decomposition analysis \cite{naturespreader}. Java et al. compared some of network properties of Twitter using users' profiles from different continents \cite{javawhywetwitter}. They found users in Europe and Asia tend to have higher reciprocity and clustering coefficient values in their corresponding sub graphs.

For other online social networks, Flickr, LiveJournal, Orkut, and YouTube are studied in \cite{mislovemeasurement}. Ahn et al. researched on the topological characteristics of Cyworld, MySpace, and orkut \cite{ahnanalysisoftopological}. They examined average degree, average clustering coefficient, assortativity, degree of separation, and other properties of these online social network services.

Micro-blogging services in China experienced rapid growth. It is believed that Sina Weibo, the most popular one in China, may exceed Twitter in the number of users due to the huge Chinese netizen base. However, there are few quantitative works on Sina Weibo and the difference between these two micro-blogging magnates. Gao et al. studied users' basic behaviors such as access ways, writing style, topics, and interest change. They analyzed more than 40 million micro-blogging activities but didn't involve following relations \cite{comparativesina}. Yin et al. studied the patterns of advertisement propagation in Sina Weibo \cite{adsina}. They extracted propagation features such as volume, topology, and time then used K-means clustering algorithm to group the messages.

Our work of following preference is also related with link analysis. Chen et al. studied friend recommendations designed to help users find known, off-line contacts and discover new friends on social networking sites \cite{chenmakenewfriends}. Hopcroft proposed a machine learning model to study the two-way relationship prediction in social network \cite{hopcroftfollowback}. Yang et al. analyzed the structure the spammers' networks that they marked on Twitter and found following preference inside the spammers' networks \cite{yanganalyzingspammer}. They found the criminal accounts tend to form a small-world network and the criminal hubs prefer to follow criminal accounts. Ghosh et al. found the link farming strategy that spammers use is begun with following social capitalists who are popular and prefer to follow back anyone who connects to them \cite{understandingcombating}.

\section{BASIC ANALYSIS}
\label{chap:basic}
In this section, some basic analysis is performed to Sina Weibo to study its network topology and other features before delving into the following preference analysis, since Sina Weibo hasn't been widely studied as Twitter. The results are compared with those of Twitter.

\subsection{Degree Distribution}
\label{chap:degreedistri}
\begin{figure}[t]
\centering
\subfigure[Distributions of followers]
{
\epsfig{file=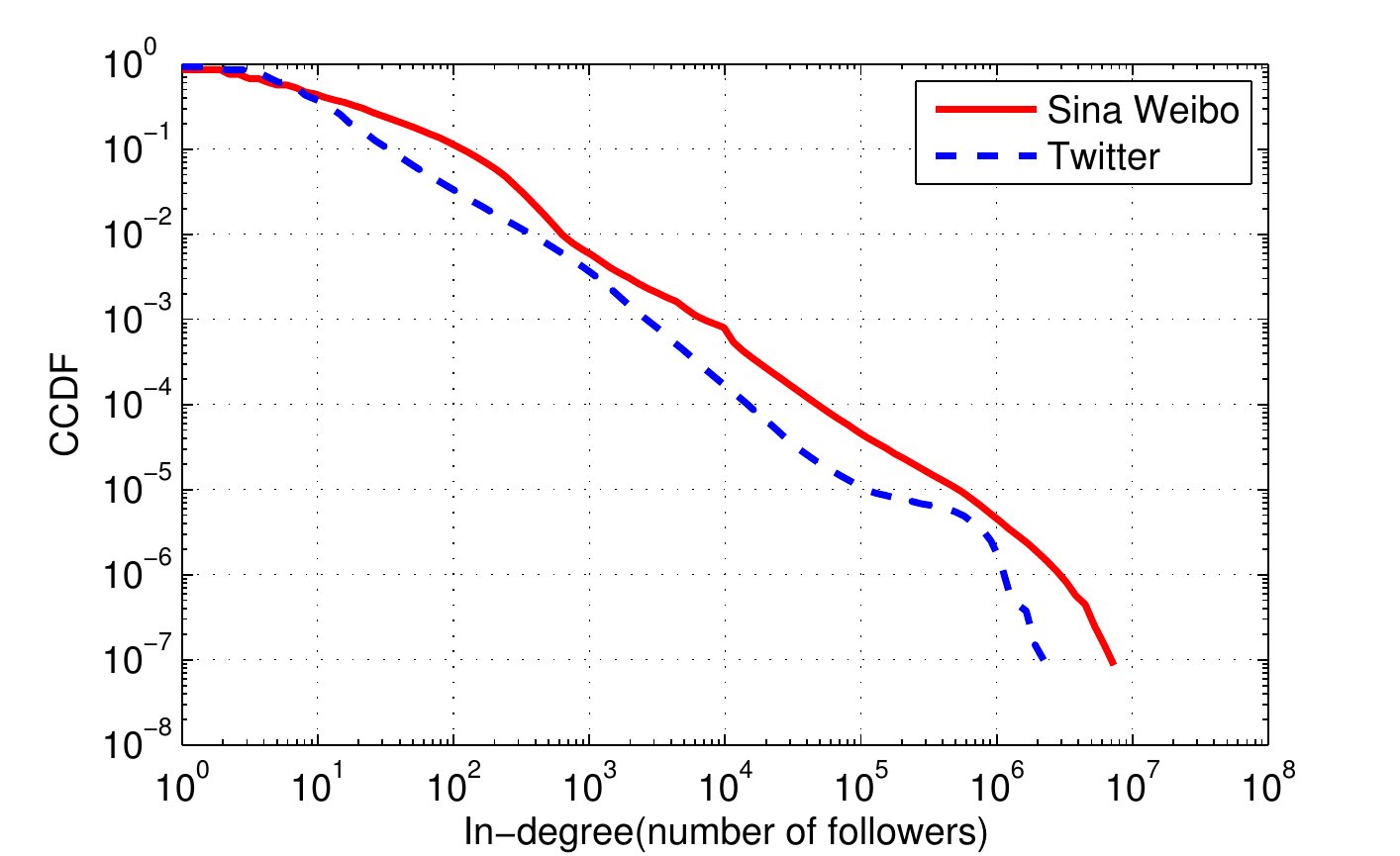, width=2.5in}
\label{indegreedistribution}
}
\subfigure[Distributions of followings]
{
\epsfig{file=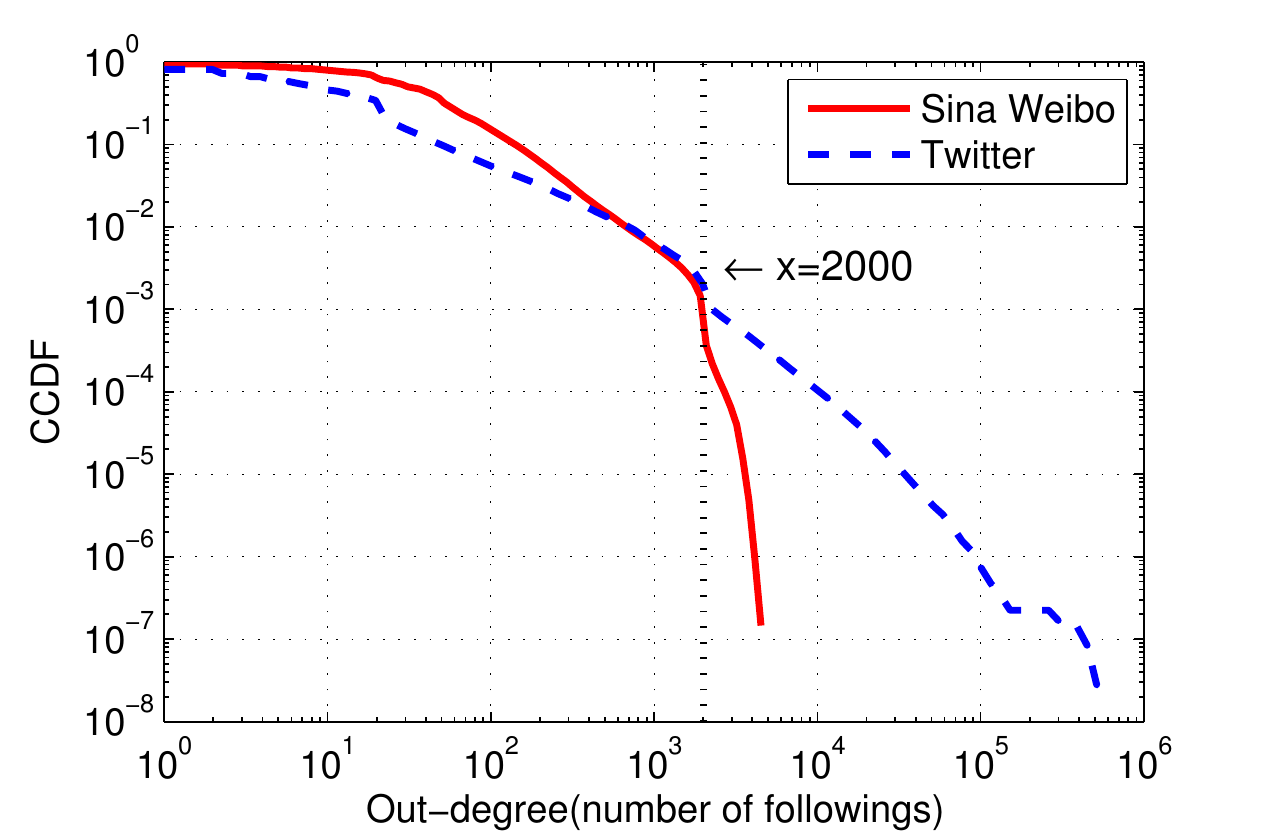, width=2.5in}
\label{outdegreedistribution}
}
\caption{Distributions of in/out-degree (followers/followings) of Sina Weibo and Twitter.}
\label{degreedistribution}
\end{figure}

The network structure of Sina Weibo and Twitter can be modeled as directed graphs. Each node has nodes linking to it and nodes it links to, corresponding to the followers and followings. Figure \ref{degreedistribution} displays the distributions of in-degree (followers) and out-degree (followings) of Sina Weibo and Twitter, respectively. The x-axis represents the number of followers or followings and the y-axis represents complementary cumulative distribution function (CCDF).

There are some glitches in Figure \ref{degreedistribution}. Firstly, in Figure \ref{outdegreedistribution}, there are a rapid decrease of the solid line and a slight but noticeable decrease of the dashed line, where the number of followings is around 2,000. Sina Weibo limits the maximum number of followings and only few VIP members can break the upper bound. The limit was also available for Twitter but removed after 2009.

Secondly, the tails of dashed lines represent the online celebrities such as actors/actresses, TV show hosts, musicians or singers, and news media. Besides, the proportion of them is higher than power-law distribution predicts in Twitter but this characteristic is not found in Sina Weibo. It reflects the global coverage property of Twitter, where world celebrities gather, while Sina Weibo is locally used.

Thirdly, there exists a gap between solid line and dashed line in Figure \ref{indegreedistribution}. The solid line is above the dashed one, where the number of followers is larger than 10. It indicates that Twitter has more users with fewer than 10 followers than Sina Weibo.

In Figure \ref{indegreedistribution}, both the solid and dashed lines approximately fit the power-law distribution. The power-law coefficient of the solid line is 2.3336 and that of the dashed line is 2.1363. Many previous researches reported that most social networks have a power-law distribution. The in-degree power-law coefficient of Twitter is reported as 2.276 in \cite{koreatwitter} and 2.4 in \cite{javawhywetwitter}. Mislove et al. reported the results for other social networks in \cite{mislovemeasurement}. The in-degree power-law coefficient of Flickr is 1.78, that of LiveJournal is 1.65, and that of YouTube is 1.99.

\subsection{Activeness}
\begin{figure}[t]
\centering
\epsfig{file=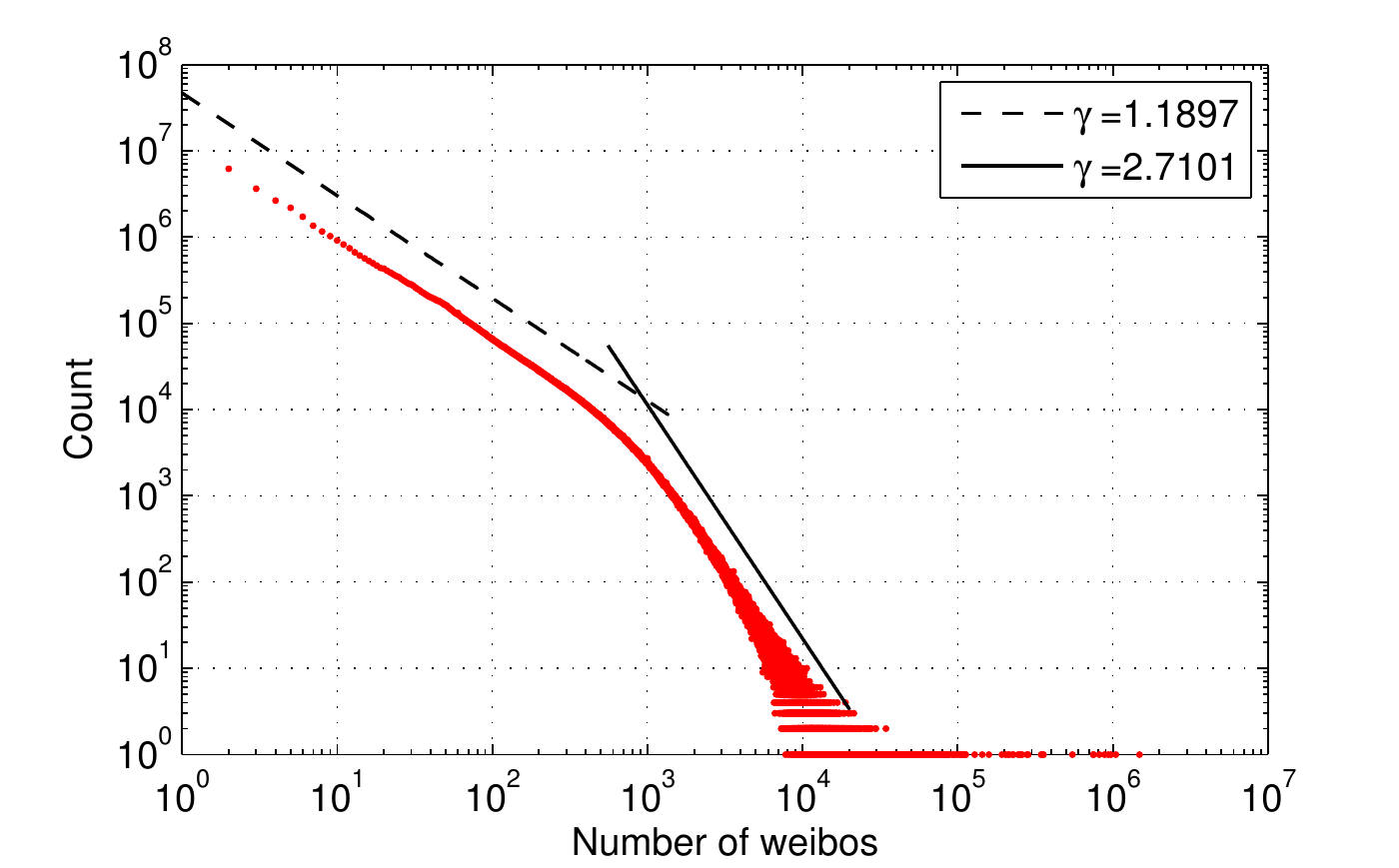,width=2.7in}
\caption{Distribution of weibos of Sina Weibo.}
\label{weibodistribution}
\end{figure}

The number of micro-blogs, which are called ``tweets'' on Twitter and ``weibos'' on Sina Weibo, is a measure of activeness. In fact how many followers one user has can't measure the activeness: a pop star updating his Sina Weibo occasionally is obviously less active than a wordy normal person posting tens of weibos every day; however the pop star has much more followers. Figure \ref{weibodistribution} displays the distribution of weibos, which is a two-stage power-law distribution with heavy tails. The curvefits a power-law distribution with exponent of 1.1897, where the number of weibos is fewer than $1,000$. The curve fits a power-law distribution with exponent of 2.7101, where the number of weibos is in the range of $1,000$ to $10,000$ . Heavy tails represent a very small number of users who have posted tens of thousands of weibos.

 As the number of weibos is a measure of user's activeness, the two-stage power-law distribution in Figure \ref{weibodistribution} shows that activeness is distributed differently from followers and followings. Please recall the power-law coefficient in Figure \ref{indegreedistribution} is 2.3336. At the first stage of the distribution, the weibos' power-law coefficient is smaller than that of followers and followings, which indicates activeness is easier to accumulate at low level. While at the second stage, the coefficient becomes larger, which indicates that it becomes hard to maintain activeness at high level.

\begin{figure}[t]
\centering
\subfigure[Weibos with followers]
{
\epsfig{file=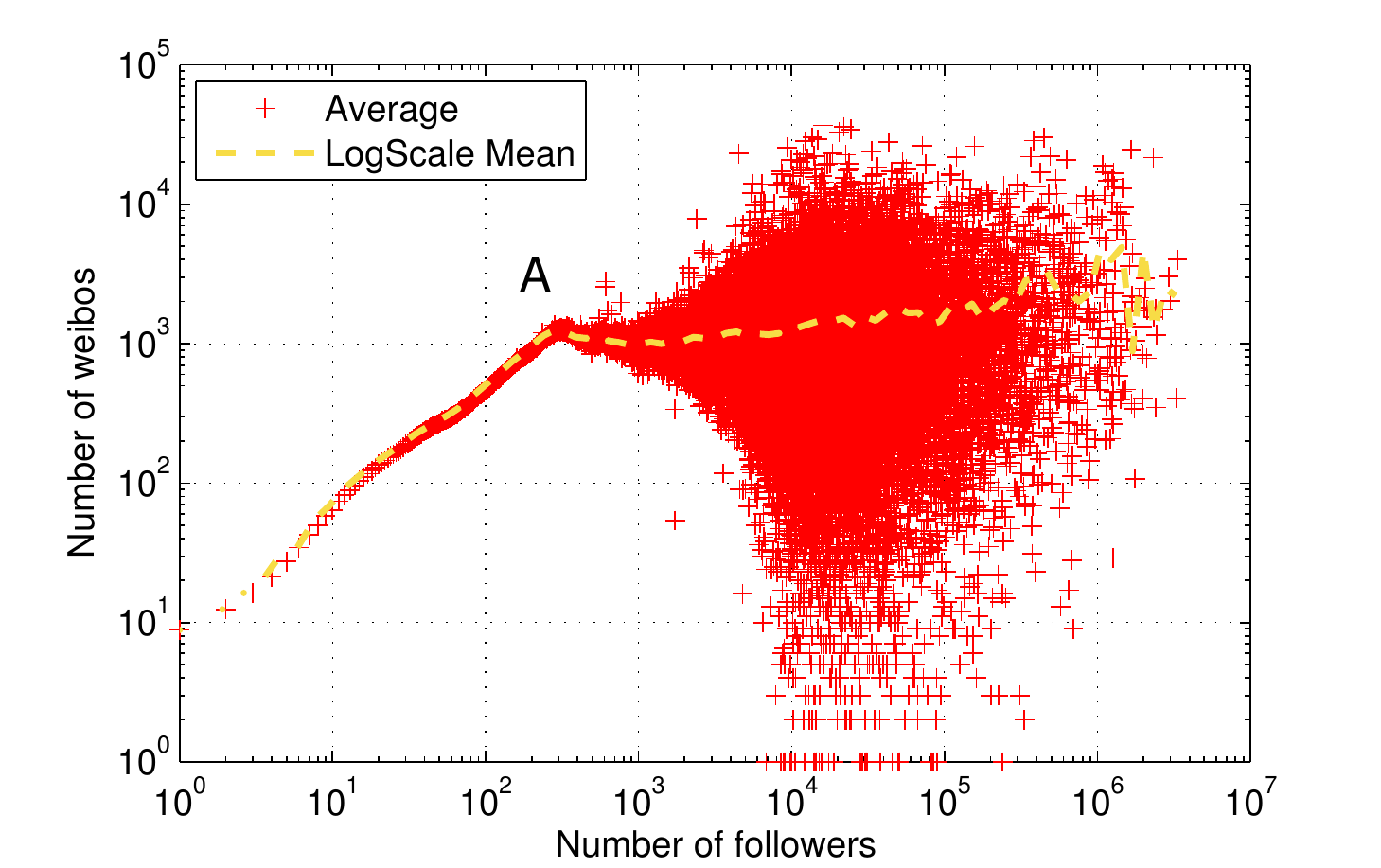,width=3in}
\label{weibofollower}
}
\hspace{1in}
\subfigure[Weibos with followings]
{
\epsfig{file=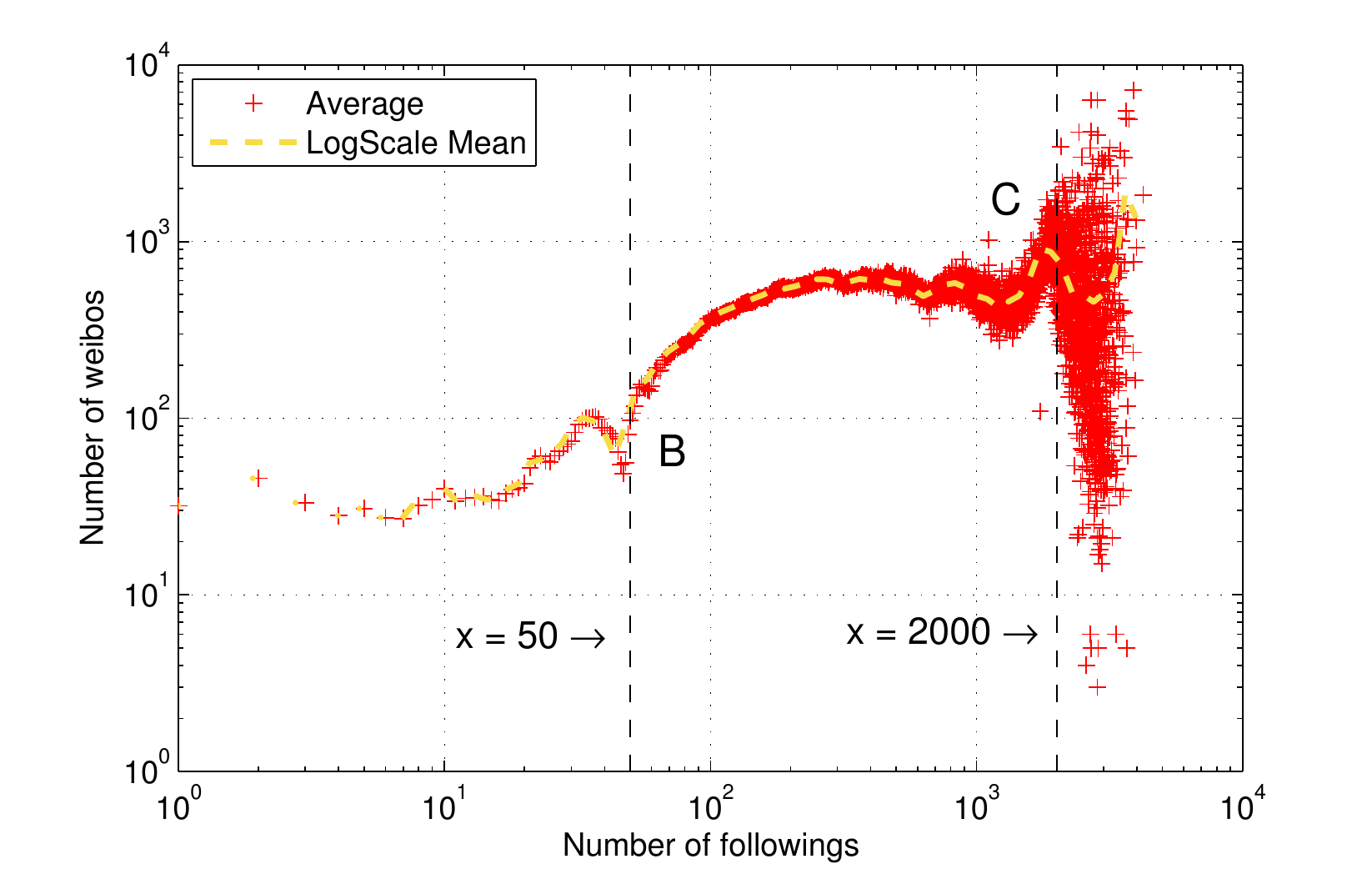,width=3in}
\label{weibofollowing}
}
\caption{The number of weibos and that of followers/followings of Sina Weibo and Twitter.}
\label{weibofollowerfollowing}
\end{figure}

To gauge the statistical correlation between them, the number of weibos (y-axis) against that of followers (x-axis) is plotted in Figure \ref{weibofollower}. Figure \ref{weibofollower} shows a positive correlation but ``+'' disperses when the number of followers increases. In Figure \ref{weibofollower}, the dashed line has an inflection at the point A. Before that point, the number of weibos that a user posts is around 7 times the number of followers he has. The correlation becomes weaker beyond the point A. However, the mean of the ``+'' in log scale still keeps a slow growth.

Besides, the correlation between users' weibos and followings is plotted in Figure \ref{weibofollowing}. The irregularity around the point B is because of the recommended system. As soon as a new user registers, Sina Weibo will recommend a set of users for him to follow which results in many users carelessly have around 50 followings initially. The cut-off around the point C is due to the upper bound limit of followings that is also observed in Figure \ref{degreedistribution}. Compared to Figure \ref{weibofollower}, the correlation between activeness and the number of followings shows similar features. It is reported that the correlation between the number of followers/followings and the number of tweets also shows a positive trend in Twitter \cite{koreatwitter}.

\subsection{Degree of Separation}
\label{ch:degreeseparaiton}
The concept of degree of separation came from Stanley Milgram's famous ``six degree of separation'' experiment \cite{milgramsmallworld} which concludes any two people in the world can be connected by no more than six people on average. Ever since then, this experiment is tested on various networks. Watts and Strogatz proposes the ``small-world'' model in \cite{smallworld} to model networks with small degree of separation.

Two users are called friends when they follow each other on Twitter or Sina Weibo. In fact only \emph{36.2\%} of relations on Twitter and \emph{20.3\%} of relations on Sina Weibo are reciprocal. Sina Weibo has an even lower reciprocity rate, which emphasizes the social media role of Sina Weibo. The friend preference will be studied in next section. The degree of separation is studied on friends' network.

\begin{figure}[t]
\centering
\epsfig{file=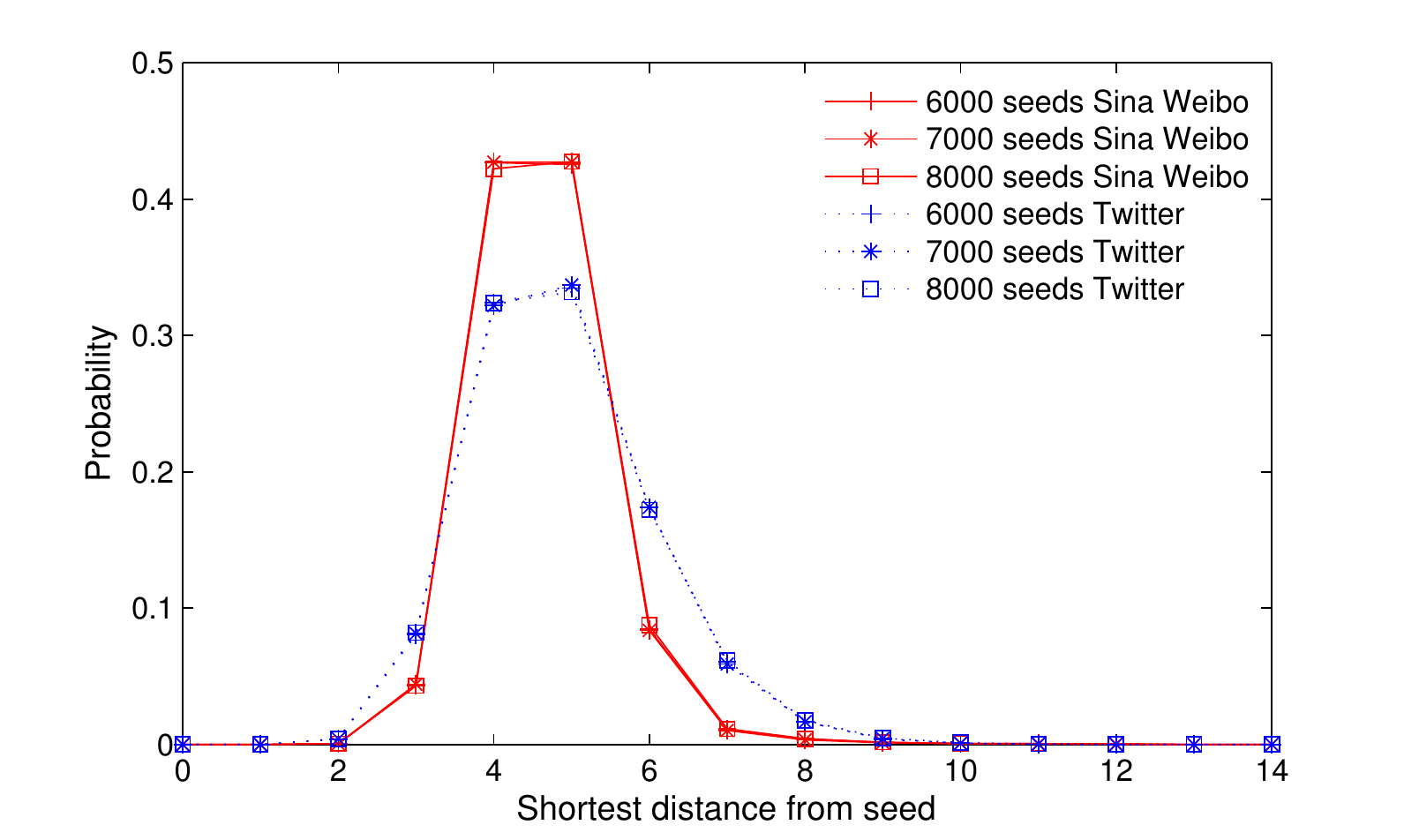,width=3in}
\caption{Degree of separation of Sina Weibo and Twitter.}
\label{fi:degreeseparation}
\end{figure}

The general method to find the shortest path length over graph is the Dijkstra algorithm. Considering the time complexity of standard Dijkstra algorithm is $O(N^3)$, where $N$ is the number of nodes, it is unacceptable for Sina Weibo and Twitter with millions users. Snowball sampling \cite{snowballsampling} is used to reduce the time complexity and to obtain an approximate result. Snowball sampling randomly picks seed nodes, performs a breadth-first search and a list of nodes marked with the distance from the seeds is obtained. Counting the nodes at each distance gives a histogram of the path length. Distributions of the shortest distance from the seed is plotted in Figure \ref{fi:degreeseparation} with $6,000$, $7,000$, and $8,000$ seeds for both Sina Weibo and Twitter. The distributions almost overlap completely as the number of seeds increases that means these seeds are enough to estimate average distance.

The average distance between arbitrary two users is $4.86$ for Twitter and $4.63$ for Sina Weibo. The effective diameter \cite{effectivediameter} of graph is defined as the 90th percentile distance and it is $5.89$ for Twitter and $5.06$ for Sina Weibo. Compared with former researches, Kwak et al. reported the average distance is $4.12$ for their data set of Twitter \cite{koreatwitter}. Other online social networks are also analyzed: the average distance is $5.67$ for Flickr, $4.25$ for Orkut, and $5.10$ for YouTube \cite{graphstructureweb} \cite{measureosn}. The average distance of Sina Weibo and Twitter is quite short for the size of them. Enough if friend relations are minor part of the total relations, the short average distance reflects the entire network is tightly connected. With a smaller degree of separation and effective diameter, it is suggested that the network structure of Sina Weibo is even tighter and more complex.

\section{FOLLOWING PREFERENCE}
\label{chap:followingpreference}
In this section, the following preference of Sina Weibo and Twitter users is analyzed. The assortative mixing, friends' similarity, following distribution, and edge balance ratio are studied to reveal the difference in structure at macro level and underlying reasons of users' behaviors.

\subsection{Attractive Features for Chinese Users}
The following preference of the followers determines users with certain features will attract more followers than others. Analyzing the attractive features helps to deduce the following preference. In our data set, each Sina Weibo user has 89.7 followers on average. Male users take up 52.46\% of the total. Each male user has 87.6 followers on average while the number for female users is 92.0. Sina Weibo introduced a certification system to reduce fake information. Users apply to the system to become verified users or ``v-user'' for showing their real identity on their homepages. Even if verified users take up only 0.34\% of the total users, each of them has 9,455.0 followers on average, in contrast to that, unverified users only have 57.5 followers, which is much fewer than the global average.

\begin{figure}[t]
\centering
\epsfig{file=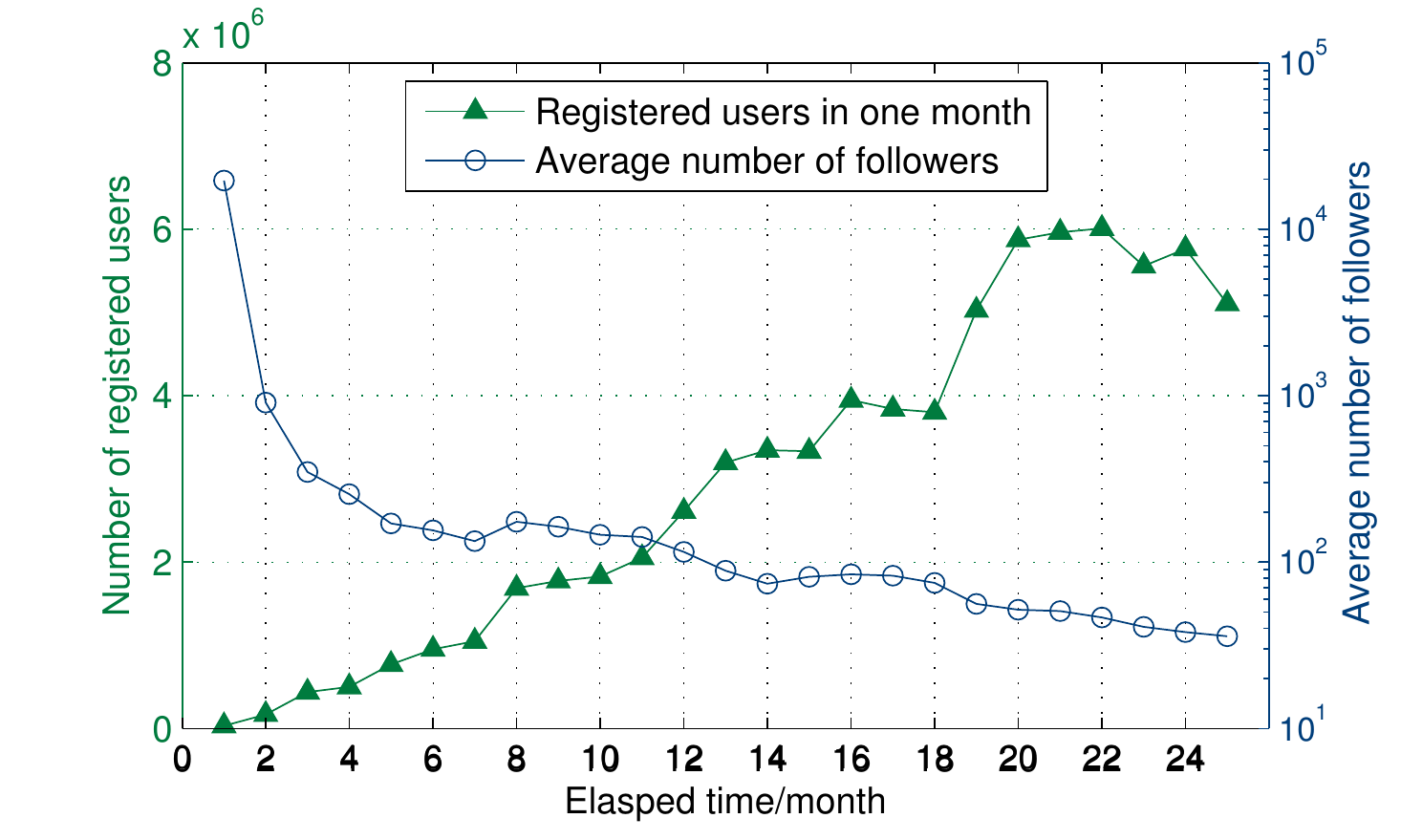,width=3.3in}
\caption{Number of average followers and registered time of Sina Weibo.}
\label{fi:createtime}
\end{figure}

The registered time of a user also affects how many followers he can attract. Figure \ref{fi:createtime} displays the average number of followers against registered time. The x-axis represents the elasped time in month since Sina Weibo opened in August, 2009. The left-side y-axis represents the number of registered users each month. The right-side y-axis represents the average number of followers of the users who register in the corresponding month. The number of registered users in each month increases over time and the earlier registered users have more followers. It should be noted that when Sina Weibo opened, most of the first group of users are celebrities and it causes the high starting point. Another reason may come from the recommendation mechanism mentioned in last section.

\subsection{Friends' Similarity}

Sociology researches \cite{homophily} reported homophily in social network. Homophily is the preference of people to associate with ones similar to themselves. In this sub section, the following preference in friends' network is analyzed. Two aspects are considered for measuring users' similarity: geographic region and fame.

Users in the same city or province might know each other off-line. In our statistics, there are 621 million pairs of friends and 47.3\% of them are in the same province. Twitter doesn't have a standard format for geographic information so it is hard to parse the users' self-written location. Time zone is used to represent the location and obtains the conclusion that Twitter users with fewer than 2,000 friends are likely to be geographically close \cite{koreatwitter}.

\begin{figure}[t]
\centering
\subfigure[Sina Weibo]
{
\epsfig{file=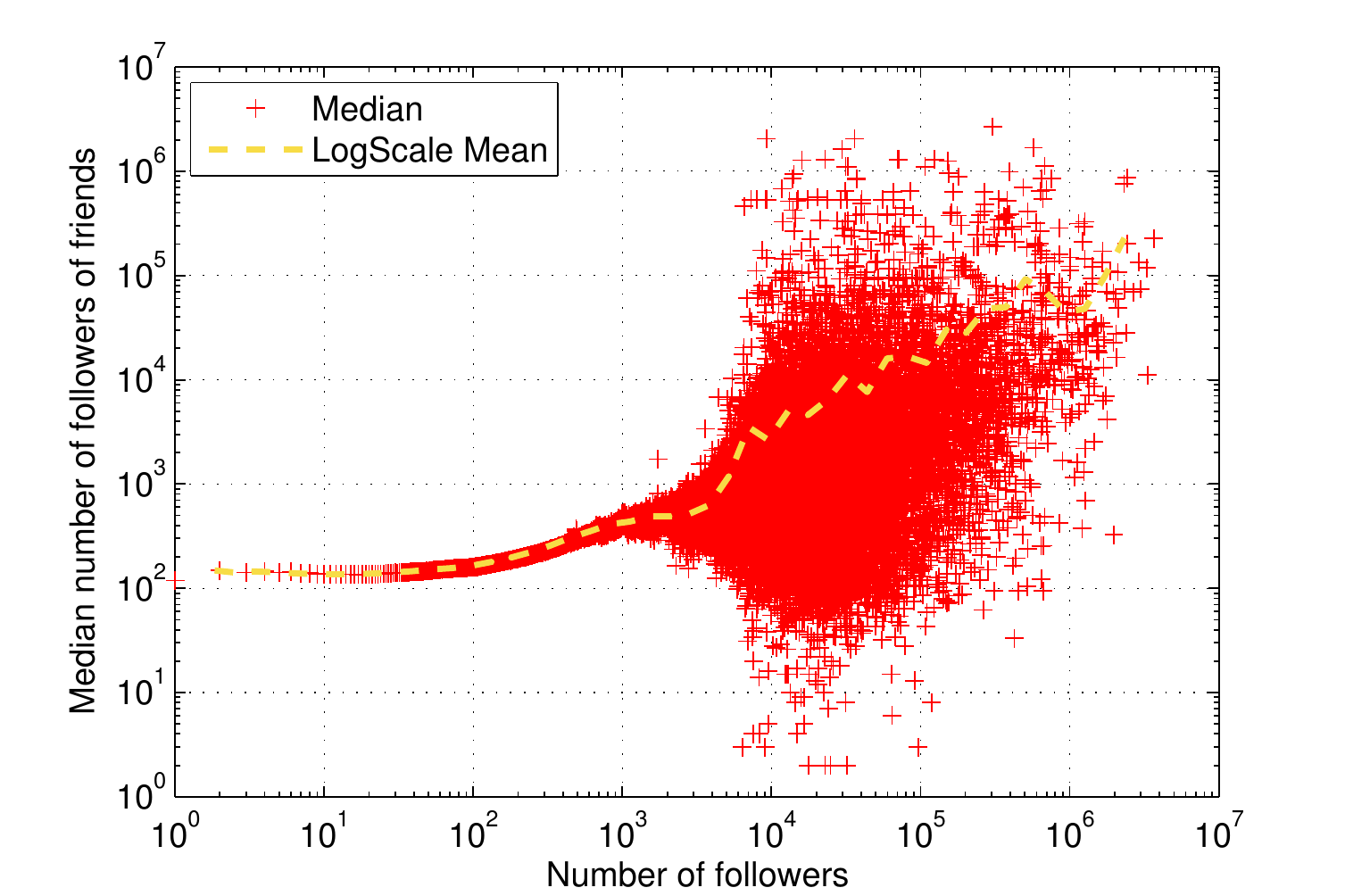,width=3in}
\label{fi:medfriendfollowerweibo}
}
\hspace{1in}
\subfigure[Twitter]
{
\epsfig{file=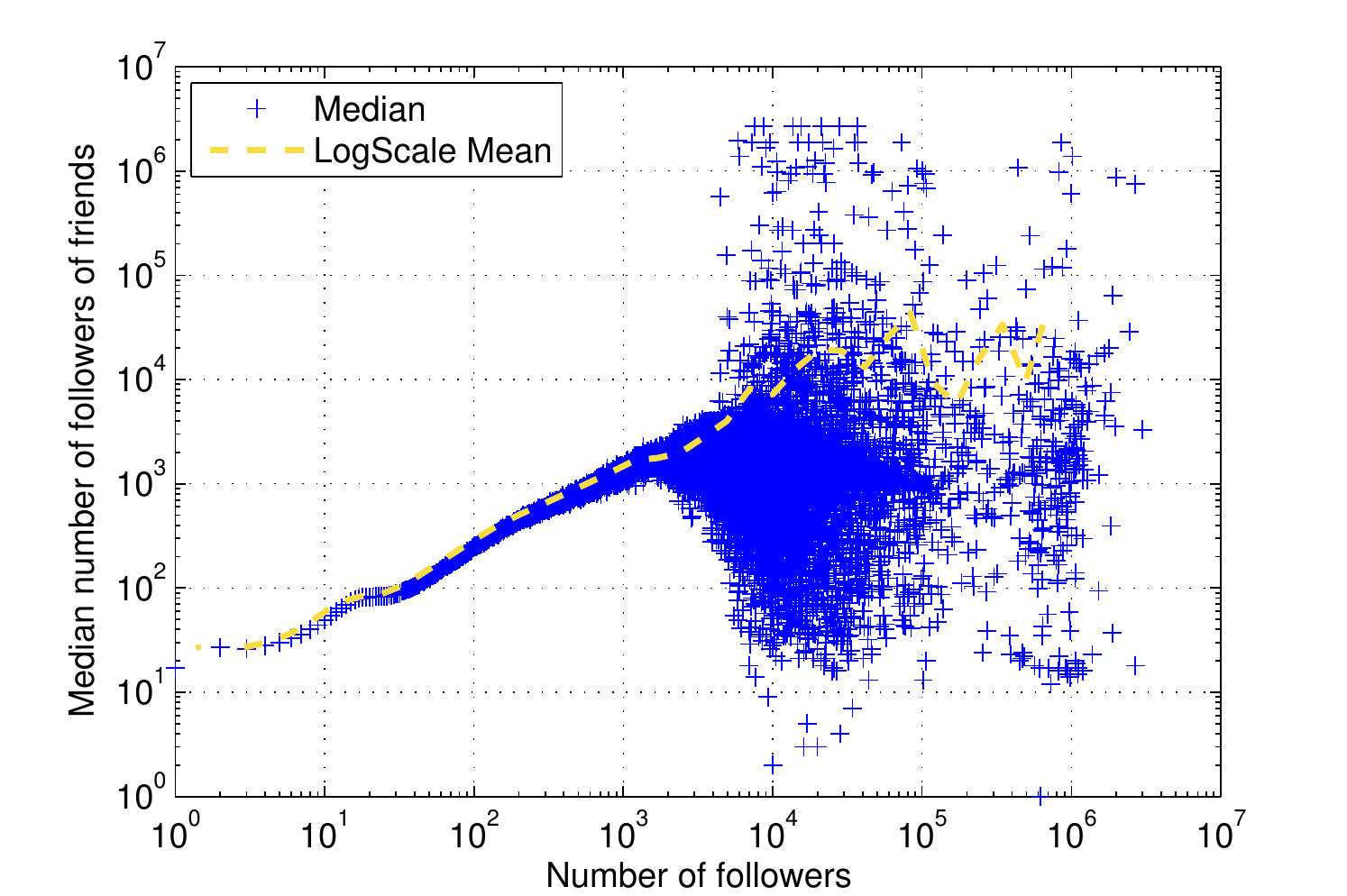,width=3in}
\label{fi:medfriendfollowertwitter}
}
\caption{Number of followers of user's friends and that of himself of Sina Weibo and Twitter.}
\label{fi:medfriendfollowers}
\end{figure}

The probability of a popular pop star following a general user is much less than following another popular star.
The number of followers is a measure of the fame. Figure \ref{fi:medfriendfollowers} plots the median of followers of a user's friends and that of himself. Every ``+'' represents the median of followers of friends over all the users with the same number of followers. The dashed line stands for the mean in log scale. In Figure \ref{fi:medfriendfollowers}, there are both significant positive correlation between the number of followers of the user's friends and that of himself when the user has fewer than 1,000 followers. Though the median numbers disperse when the number of followers become larger, the correlation is still positive when considering the mean in log scale.

The correlation between the number of followers of friends and that of a user himself is similar to degree correlation. The difference is that degree correlation usually applies for bidirectional graphs and compares the degree of a node with its neighbor nodes. The degree correlation describes in the graph a hub is more likely to connect to other hubs or less. The former situation is known as a feature of human social networks \cite{degreecorrelation}. The preference that people choose who are similar to themselves as their friends holds in Sina Weibo and Twitter in terms of graphic region and fame.

\subsection{Following Distribution}
\begin{figure}[t]
\centering
\epsfig{file=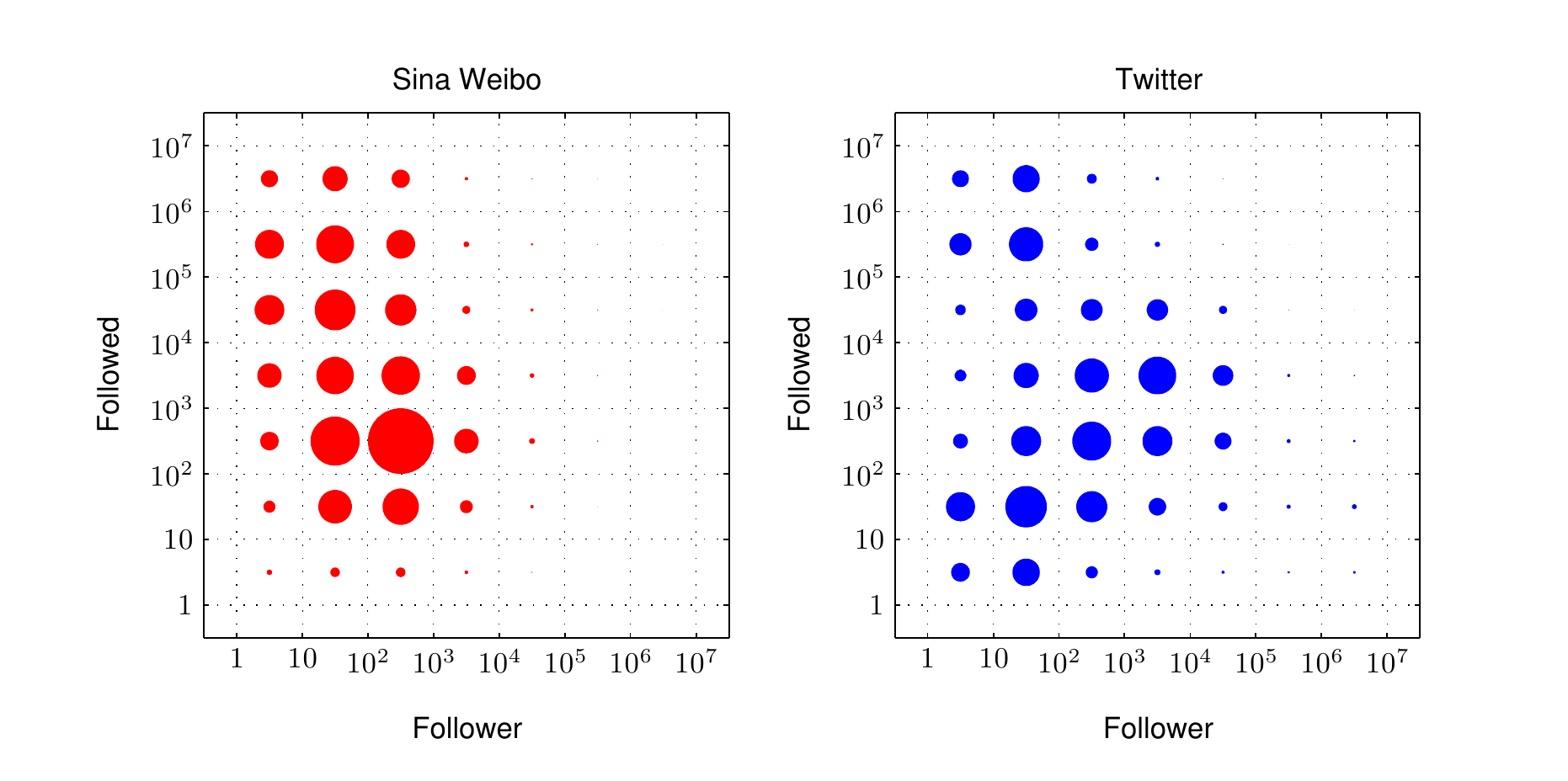,width=3.7in}
\caption{Following distribution of Sina Weibo and Twitter.}
\label{fi:followprefer}
\end{figure}

The following preference of users with different number of followers can be studied by the distribution of the following relations, which is plotted in Figure \ref{fi:followprefer}. All the users are divided into seven groups based on the number of their followers. The boundaries of these groups are shown on the axis. Every circle in Figure \ref{fi:followprefer} represents the following relations from users in the corresponding ``Follower'' group to users in the corresponding ``Followed'' group. The area of the circle represents the number of these following relations.

It is concluded that both Sina Weibo and Twitter users prefer to follow users who have the similar or more number of followers because the circles above the diagonal are larger than those below the diagonal. Besides, this kind of following preference is more significant for Sina Weibo users. However, Figure \ref{fi:followprefer} fails to show the following preference of celebrities, because the number of them is small. Other measurements will be used in next sub sections.

\subsection{Assortative Mixing}
Assortative mixing \cite{newmanassortative} or assortativity is a global measure of the preference of nodes to connect to similar nodes. For undirected networks, degree is always available as a property of node to calculate the assortative mixing by degree. For directed networks, an approach of assortativity by a set of four assortativity measures is introduced in \cite{directassortative}. Let $\alpha, \beta \in \{in,out\}$ be index of the degree type, and $s^{\alpha}$ and $t^{\beta}$ denote the in-degree or out-degree of the source node and the target node for edge $i$. The definition of assortativity is given by
\begin{equation}
r(\alpha, \beta) = \frac{\sum_{i}[(s_i^{\alpha}-\overline{s^{\alpha}})(t_i^{\beta}-\overline{t^{\beta}})]}{M\sigma^{\alpha}\sigma^{\beta}},
\end{equation}
where $M$ is the number of edges, $\overline{s^{\alpha}}$ is the average in or out degree of the source node, $\sigma^{\alpha}=\sqrt{M^{-1}\sum_{i}^{}(s_i^{\alpha}-\overline{s^{\alpha}})^2}$. $\overline{t^{\beta}}$ and $\sigma^{\beta}$ are similarly defined for target node. Nodes are more likely to link to similar nodes if $r$ is more close to $1$ and less likely if $r$ is more close to $-1$. If $r$ is close to $0$, it means no significant correlation between degrees of source and target nodes.

\begin{figure}[t]
\centering
\epsfig{file=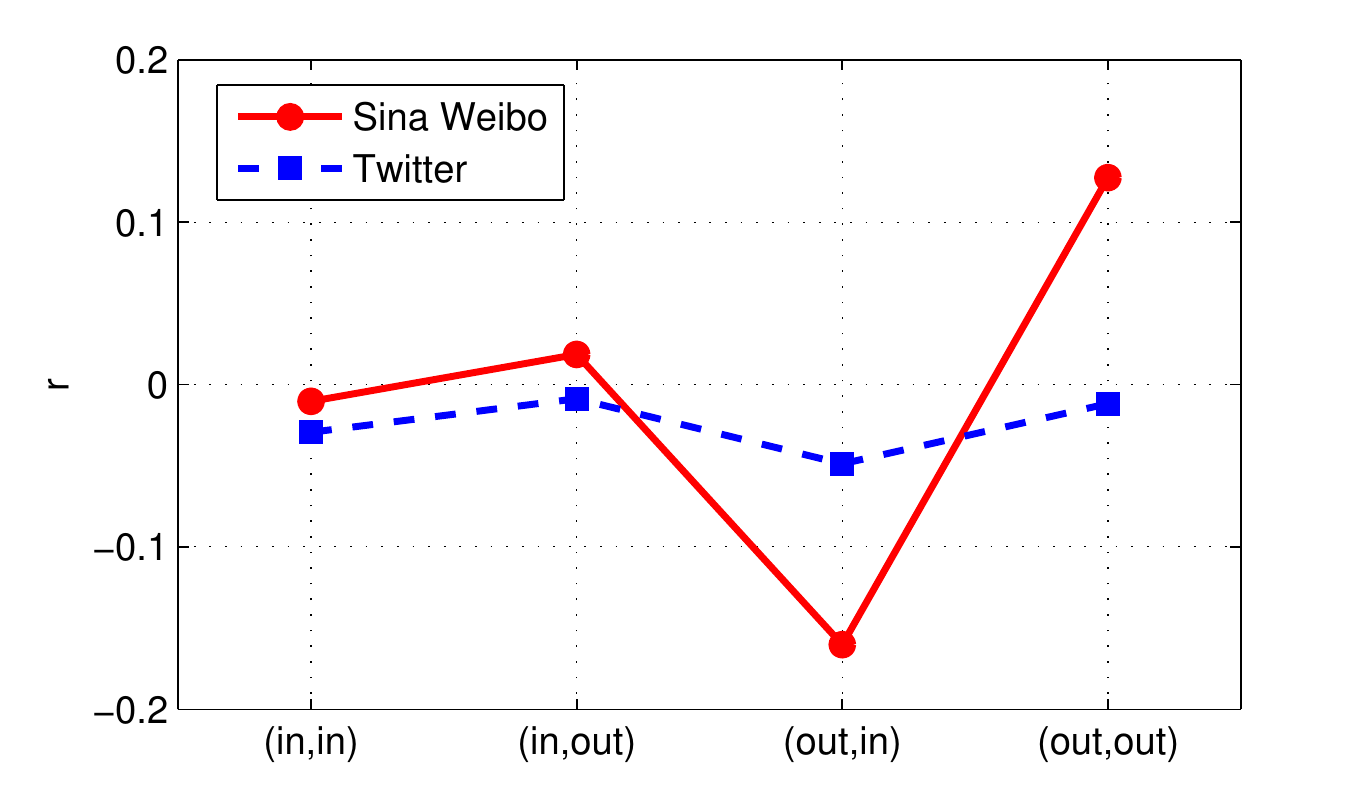,width=3in}
\caption{$r(\alpha, \beta)$ of Sina Weibo and Twitter.}
\label{fi:assortative}
\end{figure}

A set of four $r(\alpha, \beta)$ provides the profile for directed network assortativity. Such profile of Sina Weibo and Twitter relation networks are plotted in Figure \ref{fi:assortative}. It is well known that social networks usually mix assortatively \cite{newmanassortative}. However, Twitter shows slight disassortative property as the four $r(\alpha, \beta)$ are all negative and they are close to zero. While in Sina Weibo, $r(in,out)$ and $r(out,out)$ are positive and the rest are negative. The disassortative property is not hard to explain because there exists lots of ``unbalanced'' relations linked from small degree users to large degree ones, which makes the on-line network structure distinguished from traditional social networks. The remarkable difference between Sina Weibo and Twitter in Figure \ref{fi:assortative} is that $r(out,in)$ of Sina Weibo is smaller and $r(out,out)$ is larger. This indicates users with small followings tend to follow users with large followers and small followings. Therefore, normal users in Sina Weibo have stronger preference to follow people with very large number of followers.

Though assortative mixing is able to provide some macro information about the following preference, it is not detailed enough because only four scalar values are presented and assortative mixing actually make a weighted average over all the edges and is a global measurement. A new measurement, which was proposed recently, will be used to figure out more details in the next sub section.

\subsection{Edge Balance Ratio}
The unidirectional relations in Sina Weibo and Twitter results in the failure of homophily between two connected users because they can vary in many aspects such as geographic region, job occupation, influence, and fame. Edge balance ratio is a measurement to describes this balance property of a directed graph \cite{edgebalanceratio}. In directed graphs, an edge is not balanced if nodes at both ends of the edge are not equivalent in some aspects. Edge balance ratio denoted as $R$ describes the balance level of a directed edge from node A to node B and is defined as
\begin{equation}
R=\left\{
    \begin{array}{ll}
    \displaystyle{\frac{d_i({\rm B})}{d_i({\rm A})}}, &d_i({\rm A})\neq0; \\
    \infty, &d_i({\rm A})=0;
    \end{array} \right.
\end{equation}
where $d_i({\rm B})$ and $d_i({\rm A})$ are properties of node B and A such as in-degree or PageRank. Since $R$ is a property for every edge, the distribution of $R$ measures the balance property of a graph.

\begin{figure}[t]
\centering
\subfigure[In terms of the number of followers]
{
\epsfig{file=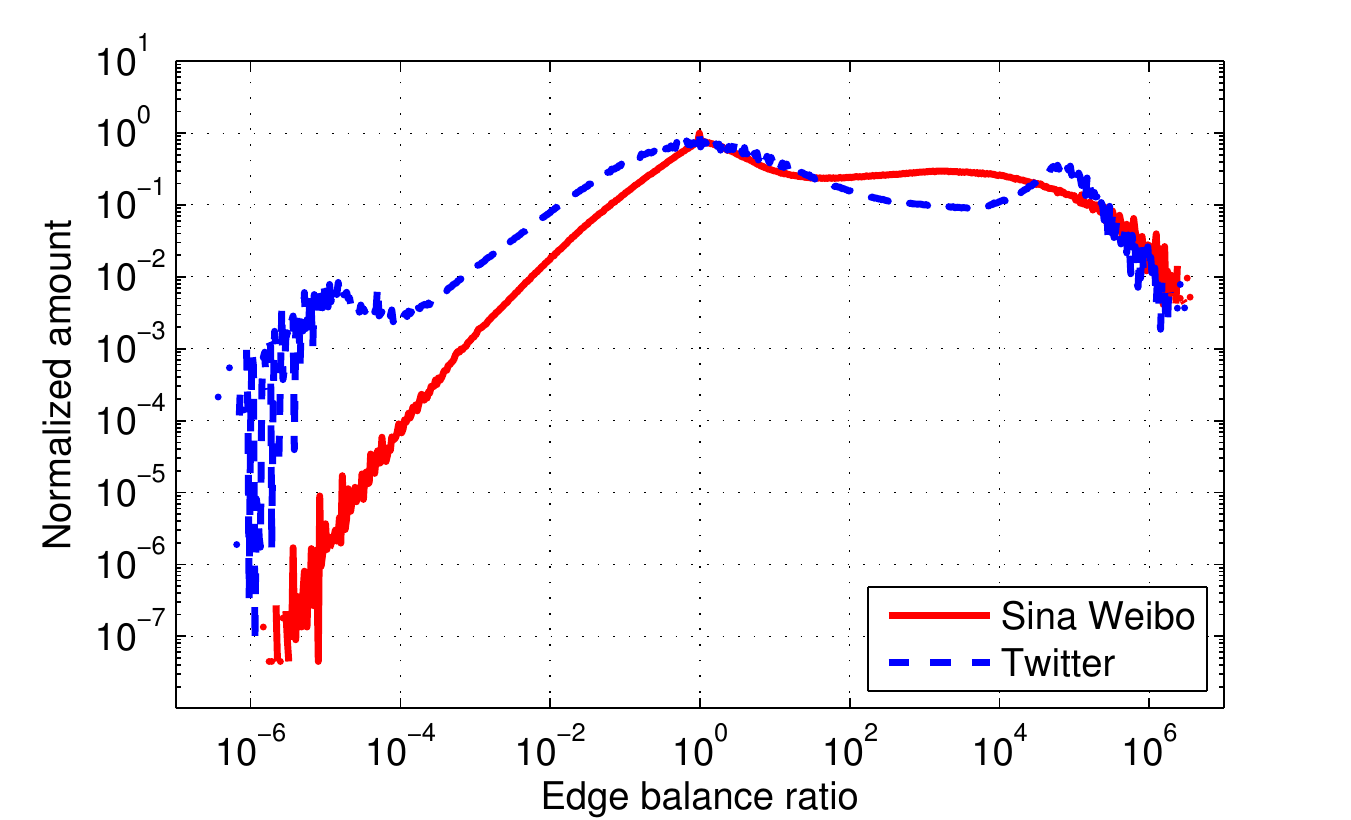,width=3in}
\label{fi:ebrfollower}
}
\subfigure[In terms of PageRank]
{
\epsfig{file=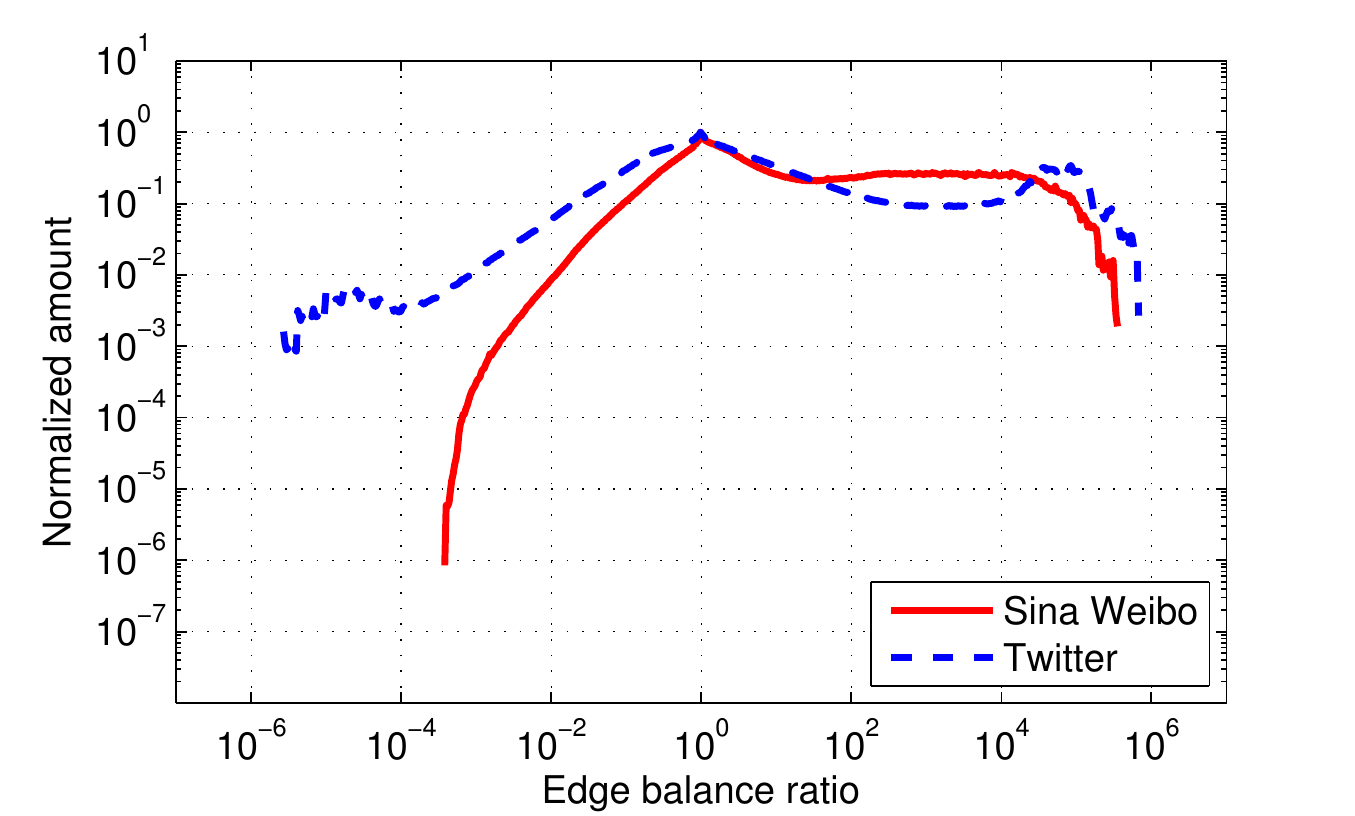,width=3in}
\label{fi:ebrpagerank}
}
\caption{Edge balance ratio of Sina Weibo and Twitter.}
\label{fi:ebr}
\end{figure}

In Sina Weibo and Twitter, the number of followers and PageRank are chosen as properties of nodes to calculate edge balance ratio. Figure \ref{fi:ebr} displays the distributions of edge balance ratio for Sina Weibo and Twitter. Every curve in Figure \ref{fi:ebr} reaches its maximum value, where $R$ equals one and its right side is higher than its left side. The dashed lines representing Twitter have two other local maximum values, where $R$ is around $10^5$ and $10^{-5}$, while the solid lines representing Sina Weibo maintain monotonicity at both sides of where $R$ equals one. The positions of solid and dashed lines indicates that Twitter has higher proportion of edges of small $R$ than Sina Weibo. Besides, Twitter has lower proportion of edges, whose $R$ is bigger than one, than Sina Weibo except for the positions, where local maximum values occur.

The edge balance ratio determines the type of relations in the network.
\begin{enumerate}[(1)]
    \item The relations with edge balance ratio far larger than one reflect users' hope to obtain news, gossips, or other type of messages from celebrities with large influence and high reputation. This is most users' important purpose to use online micro-blogging services.
    \item The relations with edge balance ratio close to one reflect users' needs to keep connections with friends\, who usually are in the same level. The homophily tells us people tend to associate with ones similar to themselves.
    \item The relations with edge balance much less than one contain rich hidden information and reveal the unique following preference.
\end{enumerate}

Figure \ref{fi:ebr} shows that in Twitter there are more the third type edges. This result indicates the network structure of Twitter might be less hierarchical than Sina Weibo, where more highly ranked users seldom follow common users.

\subsection{Summary}
The following preference is summarized as follows:
\begin{itemize}
    \item If only friends' relations are considered, both Sina Weibo and Twitter exhibit some level of homophily.
    \item Both Sina Weibo and Twitter users prefer to follow users who have the similar or more number of followers.
    \item Both Sina Weibo and Twitter have disassortative proprety, besides, Sina Weibo is more disassortative than Twitter.
    \item There are more the third type of edges in Twitter than Sina Weibo. The network structure of Twitter might be less hierarchical than Sina Weibo.
\end{itemize}

It is suggested that these results come from the different personalities and social moralities for Chinese Sina Weibo users and Twitter users. Chinese are more prudent when choosing the people to follow and they are more hierarchical and think it is inappropriate with their status to follow people with low social level. While Twitter users are more open to follow different kinds and levels of people.

\section{RANKING USERS}
\label{chap:ranking}
In this section, users are ranked by the number of followers and PageRank. Ranking gives a clear description about how important a user is. The ranking correlation reflects the following preference of top users.

\begin{figure*}
\centering
\epsfig{file=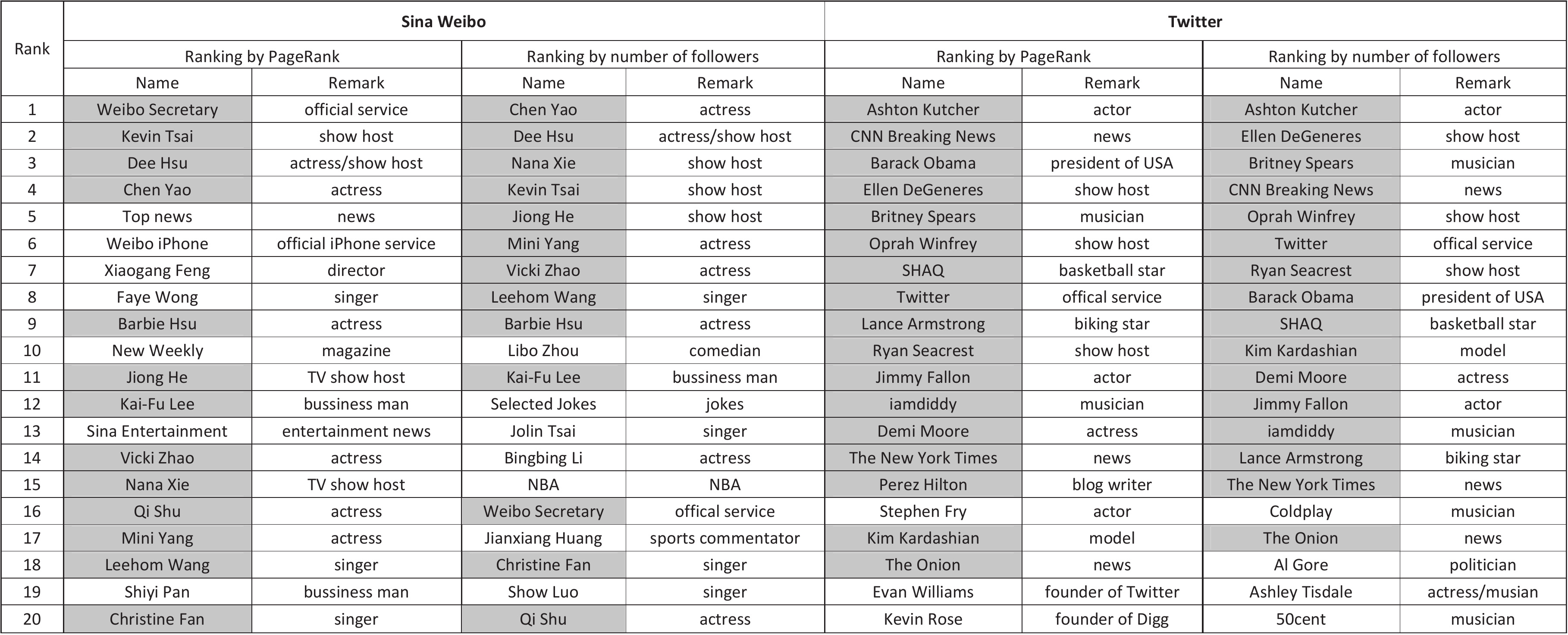,width=6.5in}
\caption{Top 20 users ranked by PageRank and the number of followers of Sina Weibo and Twitter.}
\label{fi:top20}
\end{figure*}

\subsection{Number of Followers v.s. PageRank}
Ranking users by the number of followers is a simple and directed method but it can't reflect one's influence. PageRank is an algorithm proposed to rank web pages \cite{pagerank}. It certainly applies for all directed graphs. PageRank doesn't only count the links to a web page but also evaluate importance of the linked in web page. The basic idea about PageRank is pushing the influence propagating through the links and flowing to the most influent nodes.

Figure \ref{fi:top20} shows the top 20 list ranked by the number of followers and PageRank for Sina Weibo and Twitter. Two lists for both Sina Weibo and Twitter are not exactly the same but both share many users whose names are marked in gray. In the lists of ranking by the number of followers, we find that actors, actresses, show hosts, and singers occupy most positions of the lists. In lists of ranking by PageRank, some services, news media, and politicians get higher ranks, for instance, the official service and official iPhone client service for Sina Weibo, CNN breaking news, and Barack Obama. Official services, news media, and the president of U.S. are more influential since people with many followers also have the preference to follow them thus the result is reasonable.

\subsection{Ranking Correlation}
Top 20 lists are not enough to measure the correlation between these two rankings. In this sub section, the ranking correlation is studied using generalized Kendall's tau \cite{kendalltau}. If denoting two ranking top $k$ list as $\tau_1$ and $\tau_2$, $i$ and $j$ are elements in $\tau_1$ or $\tau_2$, the ``optimistic approach'' to Kendall's tau is defined as
\begin{equation}
K^{(0)}(\tau_1,\tau_2)=\sum_{{i,j}\in \tau_1 \cup \tau_2}^{}\overline{K}^{(0)}_{i,j}(\tau_1,\tau_2).
\end{equation}

The value of $\overline{K}^{(0)}_{i,j}(\tau_1,\tau_2)$ is divided into three categories.
\begin{enumerate}[(1)]
\item   $i$ and $j$ are in both lists. If $i$ and $j$ are in the same order, $\overline{K}^{(0)}_{i,j}(\tau_1,\tau_2)=0$; otherwise $\overline{K}^{(0)}_{i,j}(\tau_1,\tau_2)=1$.
\item   Both $i$ and $j$ appear in one list and only $i$ appears in the other list. If $i$ is ranked higher than $j$, $\overline{K}^{(0)}_{i,j}(\tau_1,\tau_2)=0$; otherwise $\overline{K}^{(0)}_{i,j}(\tau_1,\tau_2)=1$.
\item   $i$ and $j$ are in different lists, let $\overline{K}^{(0)}_{i,j}(\tau_1,\tau_2)=1$.
\end{enumerate}

\begin{figure}[t]
\centering
\epsfig{file=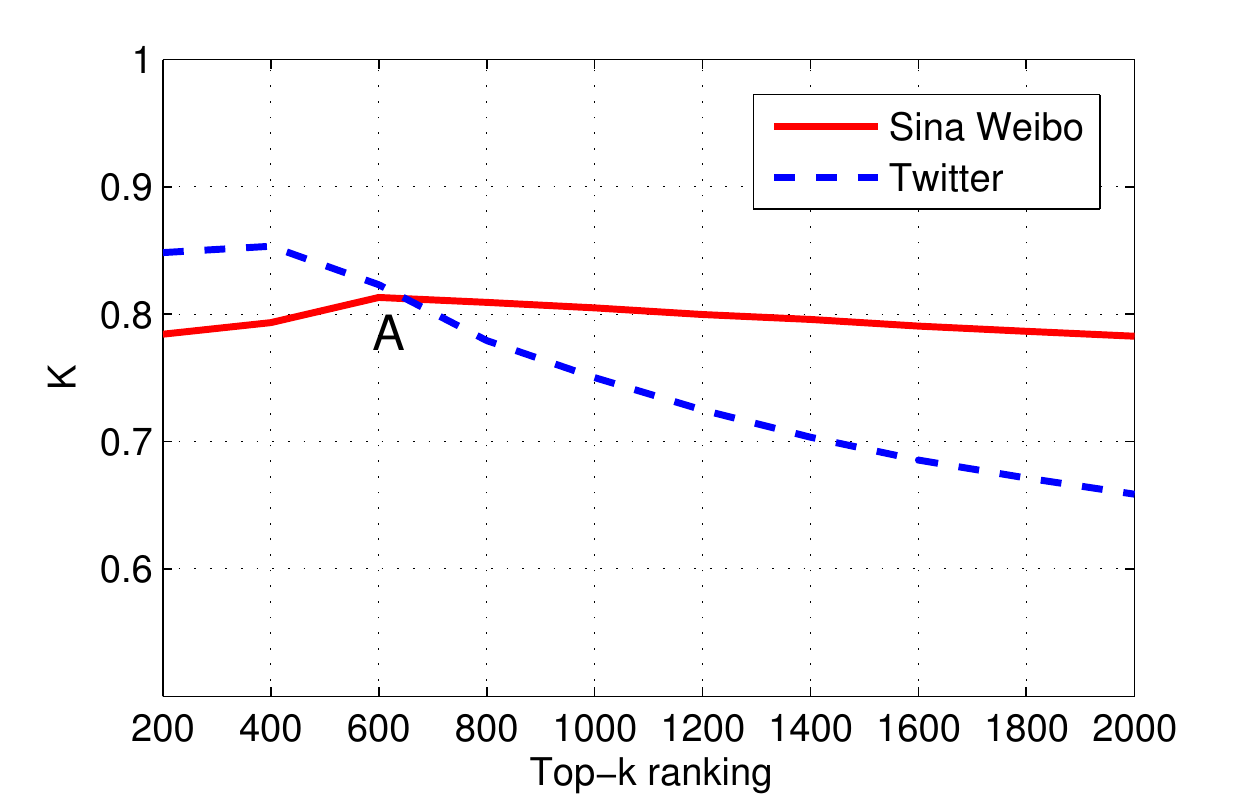,width=3in}
\caption{Ranking correlation of Sina Weibo and Twitter.}
\label{fi:rankingcor}
\end{figure}

The normalized distance $K$ \cite{normalizedK} is used to normalize the ranking correlation. $K$ is defined as
\begin{equation}
K = 1-\frac{K^{(0)}(\tau_1,\tau_2)}{k^2},
\end{equation}
where $k$ is the length of the lists. Figure \ref{fi:rankingcor} shows the correlation between ranking by number of followers and ranking by PageRank for Sina Weibo and Twitter. Both the solid line and dashed line are above $0.6$ and the dashed line representing $K$ for Twitter decreases as $k$ becomes large while $K$ for Sina Weibo keeps around $0.8$.

It is shown in the last section that users prefer to follow users who have more number of followers. The ranking correlation reflects the diversity of following preference. If the diversity is big, it may happen that important users follow the users with few followers or the users with many followers only attract the following from unimportant users. In this situation, the top-$k$ ranking correlation will be close to zero. With more top users involved, the diversity of following preference in Twitter becomes larger but it bigger little in Sina Weibo. It can be concluded that the following preference of Sina Weibo users is more concentrated than that of Twitter users.

\section{PROPAGATION}
\label{chap:propagation}
In this section, the message propagation in Sina Weibo is studied. It was reported that Twitter is a social media more than a social network \cite{koreatwitter}. The low reciprocity rate and following preference suggest that Sina Weibo has the role of social media as well. Two examples are presented to display the propagation features and critical users.

\subsection{The Characteristics of Propagation of Weibos}
There are many researches about tweets propagation but still very few about weibos in Sina Weibo. As far as we know, Yin et al. studied the patterns of advertisement propagation in Sina Weibo \cite{adsina}. Yao et al. proposed a provenance model to capture connections between micro-blog messages \cite{sinaprovenance}.

Hot weibos are tracked in order to study their propagation features. It is found that most of the hot weibos propagate no more than 10 levels, namely the farthest user from the source user is apart within 10 hops. Moreover, the hot trend is also found to disappear very fast and usually most of them can only stay hot for hours and very few can last for days. However, hot weibos always show powerful ability to reach a large coverage scale in a very short time. A possible explanation is the complexity of the network structure which has a short distance between arbitrary two users. In this sub section, two examples are presented to show their propagation features. Table \ref{tb:exampleintro} shows the brief information about the examples and their themes are both about hot social issues.

\begin{table*}
\centering
\caption{Two examples of weibo propagation.}
\label{tb:exampleintro}
\begin{tabular}{|c|c|c|c|c|c|}  \hline
Example & Source user & Weibo theme & Forwarding number & Coverage & Followers\\ \hline
\begin{minipage}[0][35pt][c]{18pt}
\begin{center}
1
\end{center}
\end{minipage}
&
\begin{minipage}[0][35pt][c]{120pt}
\begin{center}
A grassroots user popular with his sharp comments on social issues
\end{center}
\end{minipage}
&
\begin{minipage}[0][35pt][c]{120pt}
\begin{center}
 Comments on Chinese sailors seized by North Korea
\end{center}
\end{minipage}
&
\begin{minipage}[0][35pt][c]{20pt}
\begin{center}
17507
\end{center}
\end{minipage}
 &
\begin{minipage}[0][35pt][c]{50pt}
\begin{center}
 34.4 million
\end{center}
\end{minipage}
&
\begin{minipage}[0][35pt][c]{40pt}
\begin{center}
4 million
\end{center}
\end{minipage}
 \\  \hline
\begin{minipage}[0][35pt][c]{18pt}
\begin{center}
2
\end{center}
\end{minipage}
&
\begin{minipage}[0][35pt][c]{120pt}
\begin{center}
An actor and writer
\end{center}
\end{minipage}
&
\begin{minipage}[0][35pt][c]{120pt}
\begin{center}
Boycott a well-known milk brand because of substandard products
\end{center}
\end{minipage}
&
\begin{minipage}[0][35pt][c]{20pt}
\begin{center}
21835
\end{center}
\end{minipage}
 &
\begin{minipage}[0][35pt][c]{50pt}
\begin{center}
39.7 million
\end{center}
\end{minipage}
&
\begin{minipage}[0][35pt][c]{40pt}
\begin{center}
1 million
\end{center}
\end{minipage}
 \\ \hline
\end{tabular}
\end{table*}

\begin{figure}[t]
\centering
\epsfig{file=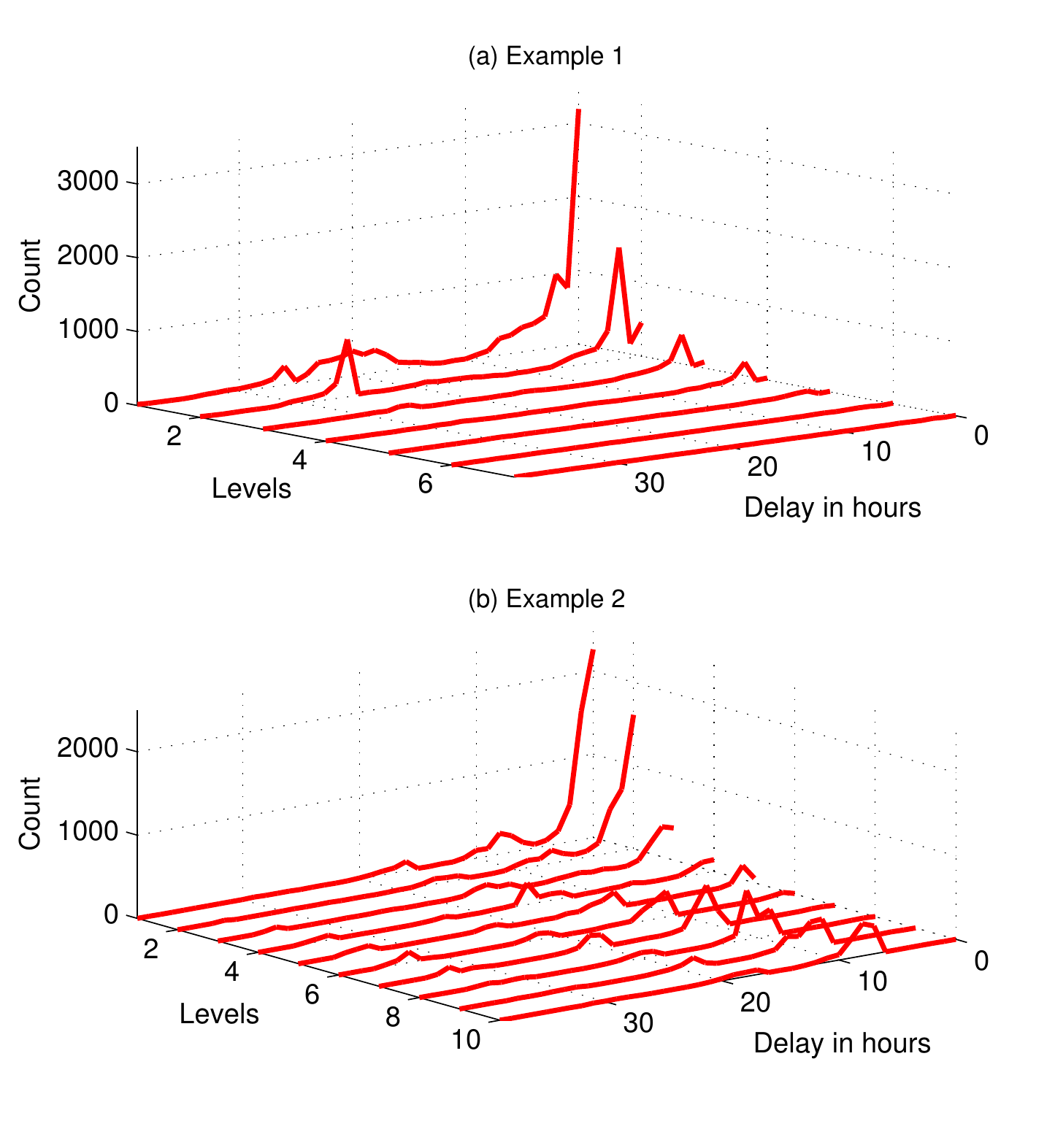,width=3in}
\caption{Distribution of propagation.}
\label{fi:leveldelay}
\end{figure}

The \emph{coverage} of a weibo is defined as the number of times people read it. Table \ref{tb:exampleintro} also displays the forwarding number and the coverage of the examples. The number of forwarded weibos is counted according to their levels and delays and it is plotted in Figure \ref{fi:leveldelay}. Both two examples cause large-scale propagation and cover large amount of users. Figure \ref{fi:leveldelay} also shows different propagation patterns. In Figure \ref{fi:leveldelay}(a), most weibos are concentrated at the position, where level is close to one. They are forwarded directly from the source weibo. While in Figure \ref{fi:leveldelay}(b) there exists some small wrinkles in the middle area. They are small propagation trends led by some participants. The propagation patterns actually related to who posts the source weibo. The source user of the first example is a grassroots commentator. But the source user of the second example is a well-known actor and writer and he is acquainted with influential people who can help him in the propagation. The subject of context here hasn't been considered but involving context will give an accurate description of the propagation pattern, which may be effective to classify the tweets or weibos.

\subsection{Critical Users in the Propagation}
In the propagation of a weibo, there are always some users who can lead relatively large secondary propagation besides the source user. They are critical users in the propagation. It is found that critical users have the ability to influence their followers to participate in the propagation no matter they are the source user or not.

Sina Weibo records the forwarding number of a weibo in the following way. Every weibo in the propagation has a forwarding number. The forwarding number of the source weibo includes the directly and indirectly forwarded weibos, while the forwarding number of others only includes the directly forwarded weibos. Assuming that user $A$ posts the source weibo $a$, user $B$ forwards weibo $a$ from $A$ and posts a forwarding weibo $b$. User $C$ forwards $b$ then $C$'s forwarding weibo, say $c$, will be counted in $B$'s and $A$'s forwarding number. But if user $D$ now forwards weibo $c$, $D$'s forwarding weibo will increase $A$'s and $C$'s forwarding number other than $B$'s. Whether a user is critical in the propagation is decided by the number of directly forwarded weibos.

\begin{figure}[t]
\centering
\epsfig{file=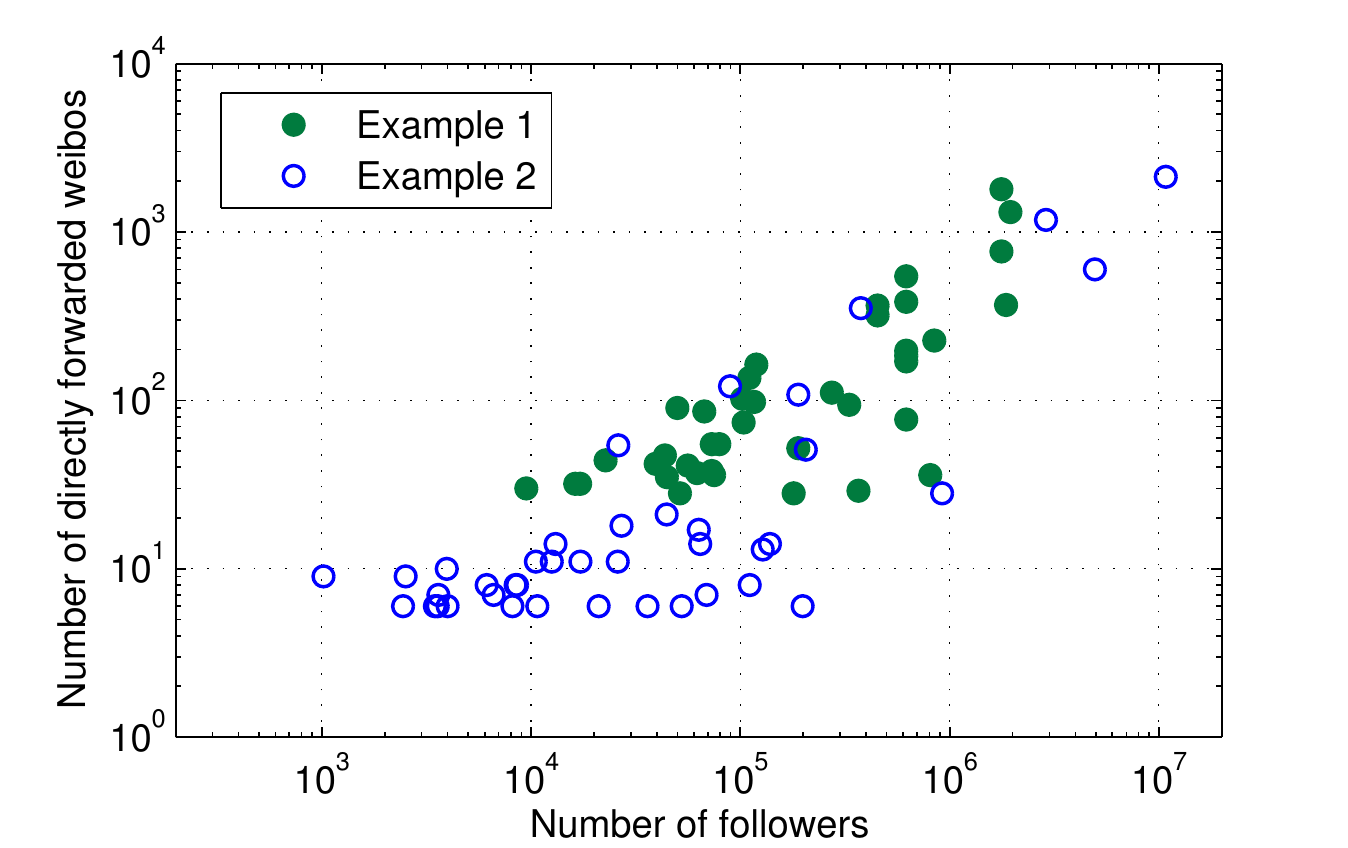,width=3in}
\caption{Critical users in propagation.}
\label{fi:criticalusers}
\end{figure}

Figure \ref{fi:criticalusers} shows the users who have more than 10 directly forwarded weibos. The positive correlation between the number of followers and the number of directly forwarded weibos indicates that users with large number of followers play critical role in the propagation even though they are not the source user.

\section{Conclusions}
\label{chap:conclusion}
This paper proposes a comparative study of users' following preference of Sina Weibo and Twitter.

\begin{itemize}
\item
Power-law degree distributions, two-stage power-law distribution of weibos, and the positive correlation between the number of followers and the number of weibos are found. The average distance and effective diameter is both very short for the size of Sina Weibo and Twitter.
\item
If only friends' relations are considered, both Sina Weibo and Twitter exhibit some level of homophily. Sina Weibo has a lower reciprocity rate, more positive balanced relations and is more disassortative than Twitter. Coinciding with Asian traditional culture, the following preference of Sina Weibo users is more concentrated and hierarchical: they are more likely to follow people at higher or the same social levels, and less likely to follow people lower than themselves. In contrary, the same kind of following preference is much weaker in Twitter, whose users are open as they follow people from various levels. The following preference derives from not only the usage habits but also underlying reasons such as personalities and social moralities, which is worthy of future research.
\item
Positive correlation between the number of followers and PageRank exists in both Sina Weibo and Twitter. Ranking correlation reflects that the following preference of Sina Weibo users is more concentrated while that of Twitter users is more diverse.
\item
Propagation levels of hot weibos are small. The hot trends disappear fast but the coverage can be quite large. It is found that the propagation patterns are related with the source users and the patterns may be effective to classify micro-blogs. It is also found that critical users, whose weibos have large forwarding number, usually have many followers.
\end{itemize}

The comparison between Sina Weibo and Twitter users' behavior provides the researchers with one excellent case to study culture differences between China and the West. This work is the first comparative study focusing on the following behavior using both large-scale data set of a global and a Chinese local online social networks.


\section{Acknowledgments}
We appreciate helpful discussions from Peng Wang and Xiangqian Che.

\bibliographystyle{abbrv}
\bibliography{sigproc}  

\end{document}